\documentclass[fleqn,usenatbib]{mnras}

\usepackage{newtxtext,newtxmath}

\usepackage[T1]{fontenc}

\usepackage[utf8]{inputenc}
\usepackage[british]{babel}

\usepackage{microtype}
\usepackage{graphicx}
\usepackage{amsmath}
\usepackage{siunitx}
\usepackage{acro}
\usepackage{csquotes}
\usepackage{cleveref}
\usepackage{xcolor}

\crefname{figure}{Figure}{Figures}
\crefname{table}{Table}{Tables}

\newcommand*{\imag}{\mathrm{i}}
\newcommand*{\euler}{\mathrm{e}}
\newcommand*{\Ms}{\ensuremath{\mathrm{M}_\odot}}
\DeclareSIUnit{\Ms}{\text{\ensuremath{\mathrm{M}_\odot}}}
\newcommand*{\textcode}[1]{\textsc{\small #1}}

\newcommand*{\email}[1]{\href{mailto:#1}{\texttt{#1}}}
\newcommand*{\q}{\enquote}

\DeclareSIUnit{\pc}{pc}
\DeclareSIUnit{\kpc}{\kilo\pc}
\DeclareSIUnit{\Mpc}{\mega\pc}
\DeclareSIUnit{\c}{\text{\ensuremath{c}}}
\DeclareSIUnit{\hHubble}{\text{\ensuremath{h}}}

\DeclareAcronym{CDM}{short=CDM, long=cold dark matter}
\DeclareAcronym{CFL}{short=CFL, long=Courant–Friedrichs–Lewy}
\DeclareAcronym{CMB}{short=CMB, long=cosmic microwave background}
\DeclareAcronym{DM}{short=DM, long=dark matter}
\DeclareAcronym{FDM}{short=FDM, long=fuzzy dark matter}
\DeclareAcronym{KDE}{short=KDE, long=kernel density estimation}
\DeclareAcronym{LSB}{short=LSB, long=low surface brightness}
\DeclareAcronym{MCMC}{short=MCMC, long=Markov Chain Monte Carlo}
\DeclareAcronym{NFW}{short=NFW, long=Navarro–Frenk–White}
\DeclareAcronym{SP}{short=SP, long=Schrödinger–Poisson}
\DeclareAcronym{UFD}{short=UFD, long=ultra-faint dwarf}
\DeclareAcronym{WIMP}{short=WIMP, long=weakly interacting massive particle}

\title[The Diversity of Core–Halo Structure in FDM]{The Diversity of Core–Halo Structure in the Fuzzy Dark Matter Model}

\author[H.\ Y.\ J.\ Chan, E.\ G.\ M.\ Ferreira, S.\ May, K.\ Hayashi, M.\ Chiba]{
Hei Yin Jowett Chan,$^{1}$\thanks{E-mail: \email{jchan@astr.tohoku.ac.jp} (HYJC),\newline \email{simon.may@mpa-garching.mpg.de} (SM)}
Elisa G.\ M.\ Ferreira,$^{2,3,4}$
Simon May,$^{2}$\footnote[1]{}
Kohei Hayashi,$^{5,6}$
Masashi Chiba$^{1}$
\\
$^{1}$Astronomical Institute, Tohoku University, Aoba-ku, Sendai, 980-8578, Japan\\
$^{2}$Max Planck Institute for Astrophysics, Karl-Schwarzschild-Straße 1, 85748 Garching, Germany\\
$^{3}$Instituto de Física, Universidade de São Paulo - C.P. 66318, CEP:
05315-970, São Paulo, Brazil\\
$^{4}$Kavli IPMU (WPI), UTIAS, The University of Tokyo, 5-1-5 Kashiwanoha, Kashiwa, Chiba 277-8583, Japan\\
$^{5}$National Institute of Technology, Ichinoseki College, Takanashi, Hagisho, Ichinoseki, Iwate, 021-8511, Japan\\
$^{6}$Institute for Cosmic Ray Research, The University of Tokyo, Chiba, 277-8582, Japan
}

\date{Accepted XXX. Received YYY; in original form ZZZ}

\pubyear{2021}

\begin{document}
\label{firstpage}
\pagerange{\pageref{firstpage}--\pageref{lastpage}}
\maketitle

\begin{abstract}
In the fuzzy dark matter (FDM) model, gravitationally collapsed objects always consist of a solitonic core located within a virialised halo.
Although various numerical simulations have confirmed that the collapsed structure can be described by a cored NFW-like density profile, there is still disagreement about the relation between the core mass and the halo mass.
To fully understand this relation, we have assembled a large sample of cored haloes based on both idealised soliton mergers and cosmological simulations with various box sizes.
We find that there exists a sizeable dispersion in the core–halo mass relation that increases with halo mass, indicating that the FDM model allows cores and haloes to coexist in diverse configurations.
We provide a new empirical equation for a core–halo mass relation with uncertainties that can encompass all previously-found relations in the dispersion, and emphasise that any observational constraints on the particle mass $m$ using a tight one-to-one core–halo mass relation should suffer from an additional uncertainty on the order of 50\,\% for halo masses $\gtrsim 10^9 \, (8\times10^{-23} \,\mathrm{eV}/(mc^2))^{3/2} \,\mathrm{M}_\odot$.
We suggest that tidal stripping may be one of the effects contributing to the scatter in the relation.
\end{abstract}



\section{Introduction}

\label{sec:introduction}

The \ac{CDM} model is one of the essential components of the standard cosmological paradigm.
In this model, \ac{DM} is described as a cold, pressureless, non-interacting fluid that dominates the matter content of the universe.
The \ac{CDM} model is extremely successful in explaining the observed large-scale structure of our universe \citep{Planck:2018vyg,BOSS:2016wmc,Pillepich:2017jle}.
However, on small scales, the behaviour of \ac{DM} is still weakly constrained and its properties are less understood.
A prominent manifestation of this is a series of possible incompatibilities found between predictions from \ac{CDM}-only simulations and observations \citep{Bullock:2017xww}.

The \ac{FDM} model is proposed to be a promising alternative to \ac{CDM} \citep[for reviews see e.\,g.][]{Hui:2016ltb,2020PrPNP.11303787N,Ferreira:2020fam,Hui:2021tkt}.
In this model, \ac{DM} is composed of ultra-light particles.
With a particle mass as light as $\SI{e-22}{\eV\per\c\squared}$, this candidate has a de Broglie wavelength of $\sim \SI{1}{\kpc}$, behaving as a wave on astrophysical scales, while on large scales it behaves like \ac{CDM}, as required by observations.
This wave behaviour on small scales leads to a series of phenomenological consequences, like the suppression of structure formation on those scales, and the formation of a core in the interior of each galaxy halo, where the field is in its ground state (soliton).
With these features, the \ac{FDM} model not only presents many predictions that can be tested using observations, but depending on its mass, it might reconcile some of the small-scale incompatibilities, like the cusp–core problem. 

The dynamics of structure formation in the \ac{FDM} model are governed by the non-relativistic Schrödinger–Poisson system of equations.
Although the computational cost of solving the coupled system in a cosmological box is known to be much more expensive than for \ac{CDM} simulations \citep{Simon2021}, \citet{Schive2014a} were able to perform cosmological \ac{FDM} simulation on an adaptive refined mesh to gain detailed insights into the non-linear structure formation.
Their self-gravitating virialised \ac{FDM} haloes are well-resolved to confirm the existence of a solitonic core at the centre of each halo, for which the density structure is approximated by the so-called soliton profile with an outer \ac{NFW}-like profile.
In addition, simulations have confirmed that \ac{FDM} indeed mimics the non-linear power spectrum of \ac{CDM} on large scales, but suppresses structure on small scales depending on the particle mass \citep{1993ApJ...416L..71W,Schive2014a,2018PhRvD..97h3519M}.

Regardless of the different numerical approaches and initial setup, several independent simulations have been performed to confirm the core–halo structure of a \ac{FDM} halo, but there is still disagreement on the relation between the core mass and the halo mass, expressed as $M_\mathrm{c} \propto M_\mathrm{h}^\alpha$ \citep{Schive2014b,Bodo2016,Mocz2017,Nori2021}. The relation depends on the mechanism of interaction between the core and the \ac{NFW} region, which is not well understood yet.
It might also depend on the formation and merger history of the haloes, as shown in \citet{Du:2016aik}.
Recent literature pointed out that the soliton is in a perturbed ground state interacting with the \ac{NFW} region, i.\,e.\ the excited states, by means of wave interference \citep{Xi2021}. 
The resulting oscillation of the soliton further complicates the analytical understanding of the relation.

The disagreement on the core–halo mass relation is of particular observational importance because many previous constraints on the particle mass of \ac{FDM} are made by dynamic modeling of dark matter-dominated galaxies, which relies on the soliton profile and core–halo mass relation predicted by simulations.
For instance, analyses of dwarf spheroidal galaxies that have often found a particle mass of $mc^2 \sim \SI{e-22}{\eV}$ or smaller \citep{Chen:2016unw,Gonzalez-Morales:2016yaf,Safarzadeh2019} are in tension with measurements like the Lyman-$\alpha$ forest measurement $mc^2 \ge \SI{e-20}{\eV}$ \citep{Rogers:2020ltq}, which constrains the \ac{FDM} mass by probing a different prediction, the suppression of structures.
For \ac{UFD} galaxies that have even smaller stellar-to-total mass ratios, some studies predicted similar particle masses as found for dwarf spheroidals \citep{Calabrese:2016hmp}, while others \citep{Safarzadeh2019} have found that the particle mass should be heavier, with the strongest bound coming from \citet{Kohei2021} with a particle mass as heavy as $mc^2 = 1.1^{+8.3}_{-0.7} \times \SI{e-19}{\eV}$ from Segue~I.
Constraints from ultra-diffuse galaxies also suggest a \ac{FDM} mass of $mc^2 \sim \SI{e-22}{\eV}$ \citep{Broadhurst2020}.
Except for the Lyman-$\alpha$ bounds, the constraints cited above depend on the assumed core–halo mass relation.
Although the origin of such incompatibilities might also be due the influence of baryons in these systems, the core–halo relation is another important aspect, and any change or uncertainty in this relation will influence the bounds on the \ac{FDM} mass cited above.

In this work, we perform new \ac{FDM} halo simulations, and use the largest cosmological \ac{FDM} simulations with full wave dynamics to date \citep{Simon2021}, to obtain a large sample of collapsed objects.
We revisit the core–halo mass relation, and find a scatter that can encompass all previously-found relations \citep[i.\,e.][]{Schive2014b,Mocz2017,Mina2020,Nori2021}.

The paper is organized as follows:
\Cref{sec:theory} reviews the equations of motion of the \ac{FDM} model in the form of the coupled Schrödinger–Poisson equations.
\Cref{sec:numerical-method} outlines the adopted numerical scheme and initial setup to perform the simulations.
\Cref{sec:results} presents the measured density profiles, scaling relations and their observational consequences.
\Cref{sec:conclusion} summarises the results and suggests a \q{to-do list} for future high-resolution simulations.

\section{Theory}
\label{sec:theory}

\subsection{The fuzzy dark matter model}

The \ac{FDM} model proposes that \ac{DM} is made of bosonic particles that are ultra-light, with a mass of $mc^2 \sim \SIrange{e-22}{e-19}{\eV}$ when all or most of the dark matter consists of \ac{FDM}.
Within this mass range, the de Broglie wavelength of this particle, given by $\lambda_\mathrm{db} \sim 1/mv$, is of the order of \si{kpc}, or slightly smaller.
This means that inside galaxies, these particles are going to behave as classical waves.
This model only has one free parameter, the particle mass $m$.
For heavier particle masses, the de Broglie wavelength (and thus the wave behaviour) would be relegated to smaller and smaller scales, so that the particles would eventually behave very closely to \ac{CDM} \citep{1993ApJ...416L..71W,2018PhRvD..97h3519M,2020JCAP...04..003G}.

As we are interested in the dynamics of this model on small scales, \ac{FDM} can be described as a non-relativistic scalar field that obeys the \ac{SP} equations.
This can be written, in comoving coordinates, as: {%
\allowdisplaybreaks
\begin{align}
    \imag\hbar\frac{\partial \psi}{\partial t}
    &= -\frac{\hbar^2}{2ma^2} \nabla^2 \psi + \frac{m\Phi}{a} \psi \,,
    \label{eq:Sceq}
    \\
    \nabla^2\Phi &= 4\pi G m(|\psi|^2 - \langle|\psi|^2\rangle)
    ,
    \label{eq:Pceq}
\end{align}%
}%
where $a = 1/(1+z)$ is the cosmological scale factor and $\Phi$ is the gravitational potential.
Note that the \ac{SP} equations follow a scaling symmetry
\begin{equation}
    \{x,t,\rho,m,\psi\} = \{\alpha x, \beta t, \beta^{-2}\rho, \alpha \beta^{-2}m, \beta^{-1}\psi\}
    \,.
    \label{eq:scaling}
\end{equation}
Therefore, this symmetry can be used to re-scale the resulting structure of a simulation to another particle mass.\footnote{%
    Since we only want to re-scale the mass, we will fix $\beta = 1$ and only change $\alpha$ to perform the scaling.
    This $\alpha$ is unrelated to the slope of the core–halo mass relation.%
}

The complex scalar field can be written in polar coordinates, 
\begin{equation}
    \psi =\sqrt{\frac{\rho}{m}}\euler^{\imag\theta} \,,
    \label{eq:polar}
\end{equation}
where the amplitude and phase are related to the fluid comoving density and velocity
\begin{align}
    \rho = m |\psi|^2
    ,\qquad
    \mathbf{v} = \frac{\hbar}{am} \nabla\theta
    \,.
    \label{eq:fluid}
\end{align}
The above relation is called the Madelung transformation \citep{Madelung1927}.
This allows us to rewrite the system of \cref{eq:Sceq,eq:Pceq} for \ac{FDM} in a hydrodynamical form:
\begin{align}
    & \dot{\rho} + 3H\rho +\frac{1}{a} \mathbf{\nabla} \cdot \left( \rho \mathbf{v} \right) = 0\,, \\ 
    & \dot{\mathbf{v}}+H\mathbf{v}+\frac{1}{a} \left( \mathbf{v} \cdot \mathbf{\nabla}  \right) \mathbf{v} = - \frac{1}{a} \mathbf{\nabla} \Phi + \frac{1}{2a^3 m^2} \mathbf{\nabla} \left( \frac{\nabla^2 \sqrt{\rho}}{\sqrt{\rho}} \right) \,,
    \label{eq:hydro}
\end{align}
with the Hubble parameter $H = \dot{a}/a$.
These equations are the Madelung equations.

The last term of \cref{eq:hydro}, the modified Euler equation, is often called \q{quantum pressure}\footnote{This term is also called \q{quantum potential} in parts of the literature since it can be rewritten in terms of a stress tensor that has off-diagonal components, hence unlike pressure. Some also claim that this term is similar to the Bohm quantum potential (see \citet{Ferreira:2020fam} for details). Here we use the historic and most commonly used term: \q{quantum pressure}.}, which has an effect of counteracting gravity.
This term is not present in \ac{CDM} and only appears in this type of models.
From the competition between these two components, hydrostatic equilibrium is reached at a defined length scale, the Jeans wavelength, below which structures will not form.
Therefore, this model predicts a suppression of structure formation on small scales.

\subsection{Non-linear structure of the fuzzy dark matter model}

A consequence of the finite Jeans length and corresponding suppression of small-scale structure formation can be seen in the suppression of small-scale power in the power spectrum of these models, and consequently the suppression of the formation of smaller haloes.
The effect of this suppression can also be seen inside haloes, where there is a highly non-linear evolution.
The interior of each halo forms a core, where there is no further structure formation and the \ac{FDM} field is in its ground state.
A gravitationally bound object thus consists of two components in the \ac{FDM} model:
The inner part – where quantum pressure dominates – is called the core, while in the outer part, gravity dominates and structure formation can happen.
The density profile of the entire halo structure can be modeled by a cored \ac{NFW} profile
\begin{equation}
    \rho(r) =
    \begin{cases}
        \rho_{\mathrm{c}} \left[ 1+0.091\left(\frac{r}{r_{\mathrm{c}}}\right)^2 \right]^{-8}
        & \text{, for $r < r_{\mathrm{t}}$}
        \\
        \rho_{\mathrm{s}} \left[\frac{r}{r_{\mathrm{s}}}\right]^{-1} \left[1+\left(\frac{r}{r_{\mathrm{s}}}\right)\right]^{-2}
        & \text{, for $r \ge r_{\mathrm{t}}$}
    \end{cases}
    \label{eq:corenfw}
\end{equation}
with the core density
\begin{equation}
    \rho_{\mathrm{c}} = \num{1.9e9}\, a^{-1} \left( \frac{\SI{e-23}{\eV}}{mc^2} \right)^{2} \left( \frac{\si{\kpc}}{r_{\mathrm{c}}} \right)^4\,.
    \label{eq:core-density}
\end{equation}
The core density is a numerical fit to the \ac{FDM} simulations from \citet{Schive2014b}.
The scale density $\rho_\mathrm{s}$ can be obtained from the continuity condition for the density
\begin{equation}
    \frac{\rho_\mathrm{s}}{\rho_\mathrm{c}} = \left[ 1 + 0.091\left(\frac{r_\mathrm{t}}{r_\mathrm{c}}\right)^2 \right]^{-8}
    \left[\frac{r_\mathrm{t}}{r_\mathrm{s}}\right]\left[1+\left(\frac{r_\mathrm{t}}{r_\mathrm{s}}\right)\right]^{2}\,.
    \label{eq:discont}
\end{equation}
Thus, the cored \ac{NFW} profile depends on three parameters $r_{\mathrm{c}}$, $r_{\mathrm{t}}$ and $r_{\mathrm{s}}$, which denote the core, transition, and scale radius, respectively.
Previous simulations show that the core structure is well fitted by the core profile with maximum error \SI{2}{\percent} up to the transition radius $r_{\mathrm{t}} \ge 3 r_{\mathrm{c}}$ \citep{Schive2014b}.
For the outer region $r > r_{\mathrm{t}}$, the profile follows the \ac{NFW} profile.

This model, imposing only continuity of the densities, does not guarantee a smooth transition.
To do so, an extra continuity condition in the first derivative of the density must be imposed in addition to \cref{eq:discont} for the model to be both continuous and smooth.
However, the resulting transition radius for a smooth transition is $r_{\mathrm{t}} < 3r_{\mathrm{c}}$, as was shown analytically in \citep{Bernal:2017oih}, which is in disagreement with previous results from simulations \citep{Schive2014b,Mocz2017}.
In this work, we will only apply the continuity \cref{eq:discont}, and allow $r_{\mathrm{t}}$ to vary. 

In \citet{Schive2014b}, a fitting function for the core–halo mass relation was obtained:
\begin{equation}
    M_\mathrm{c} = \frac{1}{4\sqrt{a}}\left[ \left(\frac{\zeta(z)}{\zeta(0)}\right)^{1/2} \frac{M_\mathrm{h}}{M_\text{min,0}} \right]^{1/3} M_\text{min,0}
    \,,
    \label{eq:chrelation}
\end{equation}
where $M_\mathrm{c}$ and $M_\mathrm{h}$ are again the core and halo masses, and $M_\text{min,0} \sim \num{4.4e7} \big(mc^2/(\SI{e-22}{\eV})\big)^{-3/2} \Ms$, and the outer exponent $\alpha = 1/3$ represents the (logarithmic) slope of the relation $M_\mathrm{c} \propto M_\mathrm{h}^\alpha$.
In order to compare with \citet{Schive2014b}, we follow their definition of halo mass $M_\mathrm{h} = (4\pi r_\mathrm{h}^3/3) \zeta(z) \rho_\mathrm{m0}$, where $r_\mathrm{h}$ is the halo's virial radius, $\rho_\mathrm{m0}$ is the background matter density and $\zeta \sim 180$ ($350$) for $z = 0$ ($\ge 1$).

Previous studies were able to confirm the empirical density profile \cref{eq:corenfw,eq:core-density} using different simulations.
However, they disagree about the form of the core–halo mass relation, calling the validity of \cref{eq:chrelation} obtained by \citet{Schive2014b} into question.
\citet{Bodo2016} performed idealised soliton merger simulations and were unable to reproduce \cref{eq:chrelation}.
\citet{Mocz2017} used a larger halo sample with simulations of a similar setup and obtained a slope of $\alpha = 5/9$, disagreeing with \cref{eq:chrelation}.
\citet{Mina2020} found the same slope of $5/9$ using cosmological simulations with a box size of \SI{2.5}{\Mpc\per\hHubble}.
Finally, \citet{Nori2021} performed zoom-in simulations by including an external quantum pressure term in an $N$-body code, and obtained a relation with yet another value of the slope, $\alpha = 0.6$.
Such disagreement between different studies indicates that there is still a fundamental lack of understanding of the core–halo structure in the \ac{FDM} model, and also generates uncertainty in any constraints on the \ac{FDM} mass which were obtained using \cref{eq:chrelation} or similar relations.
Therefore, the main motivation of this work is to revisit and clarify the core–halo mass relation.
We will further discuss the existing discrepancies in the literature and their possible origins together with our own results in \cref{sec:results-core-halo-relation}.

\section{Numerical method}
\label{sec:numerical-method}

Previous core–halo relations are obtained from different types of simulations.
The most general way is to perform a cosmological simulation, but these simulations are often restricted to end before redshift $z = 0$ and the number of well-resolved cores is limited due to computational difficulties.
A cheaper approach is to perform non-cosmological simulations of soliton mergers.
This approach allows more control of the resolution and the final halo mass, but is at risk of simulating unrealistic haloes due to the idealised, non-cosmological initial conditions.
In this work, we analyse properties of haloes from three  different sets of simulations: 1) soliton merger simulations, 2) cosmological simulations in a small box, and 3) a high-resolution large-scale cosmological simulation.
The first two sets of simulations are performed in this work, and the last was performed by \citet{Simon2021}.
All of them used the same numerical scheme, but different initial conditions.

\subsection{Numerical scheme}
The time-dependent \ac{SP} given in \cref{eq:Sceq,eq:Pceq} are discretised on a uniform spatial grid and evolved from timestep $n$ to the next timestep using the pseudo-spectral method
\begin{equation}
    \psi_{n+1}
    \approx \euler^{K \Delta t} \mathcal{F}^{-1} \left[
    \euler^{D \Delta t} \, \mathcal{F} \left[
    \euler^{K \Delta t} \psi_{n} \right]\right]\,,
\end{equation}
where $K = -\imag m\Phi/(2\hbar a)$, $D = -\imag\hbar k^2/(2ma^2)$, and $\mathcal{F}$ denotes the Fourier transform operator \citep[see e.\,g.][]{2009ApJ...697..850W}.
This scheme is second-order accurate in time and exponentially accurate in space. Each full time integration is divided into three steps, which is similar to the symplectic leapfrog, \q{kick-drift-kick}, integrator. 
Before applying the \q{kick} operator $\euler^{K\Delta t}$, the potential $\Phi$ must be updated by solving the Poisson equation shown in \cref{eq:Pceq}.

Since the numerical method is explicit, the choice of time step must follow a \ac{CFL}-like condition.
In this case, the phases of the exponential operators must be smaller than $2\pi$:
\begin{equation}
    \Delta t < \operatorname{min}\left\{ \frac{4}{3\pi}\frac{m}{\hbar}\Delta x^2 a^2, \; 2\pi a\frac{\hbar}{m \, |\Phi_{\text{max}}|}\right\}\,,
    \label{eq:timestep}
\end{equation}
where $|\Phi_{\text{max}}|$ is the maximum value of the potential.
The scale factor for the next time step is approximated by $a_\text{next}\approx a + Ha\Delta t$, which is later used to calculate the time steps for the \q{kick} and \q{drift} operators. 

At early times, the \ac{CFL} condition is determined by the \q{drift} operator. As the gravitational potential becomes deeper at later times, the \q{kick} term begins to control the choice of time step.
For example, $\sim \SI{90}{\percent}$ of the computational time is controlled by the \q{drift} term in our simulations.
The scheme restricts this work to simulations of less massive haloes, because the core radius–halo mass relation $r_\mathrm{s} \propto M_\mathrm{h}^{-\alpha}$ implies that a higher spatial resolution is required to resolve the small core radius of a massive halo, leading to smaller time steps based on the \ac{CFL} condition $\Delta t \propto \Delta x^2$.

\subsection{Initial setup}
\begin{figure}
	\includegraphics[width=\columnwidth]{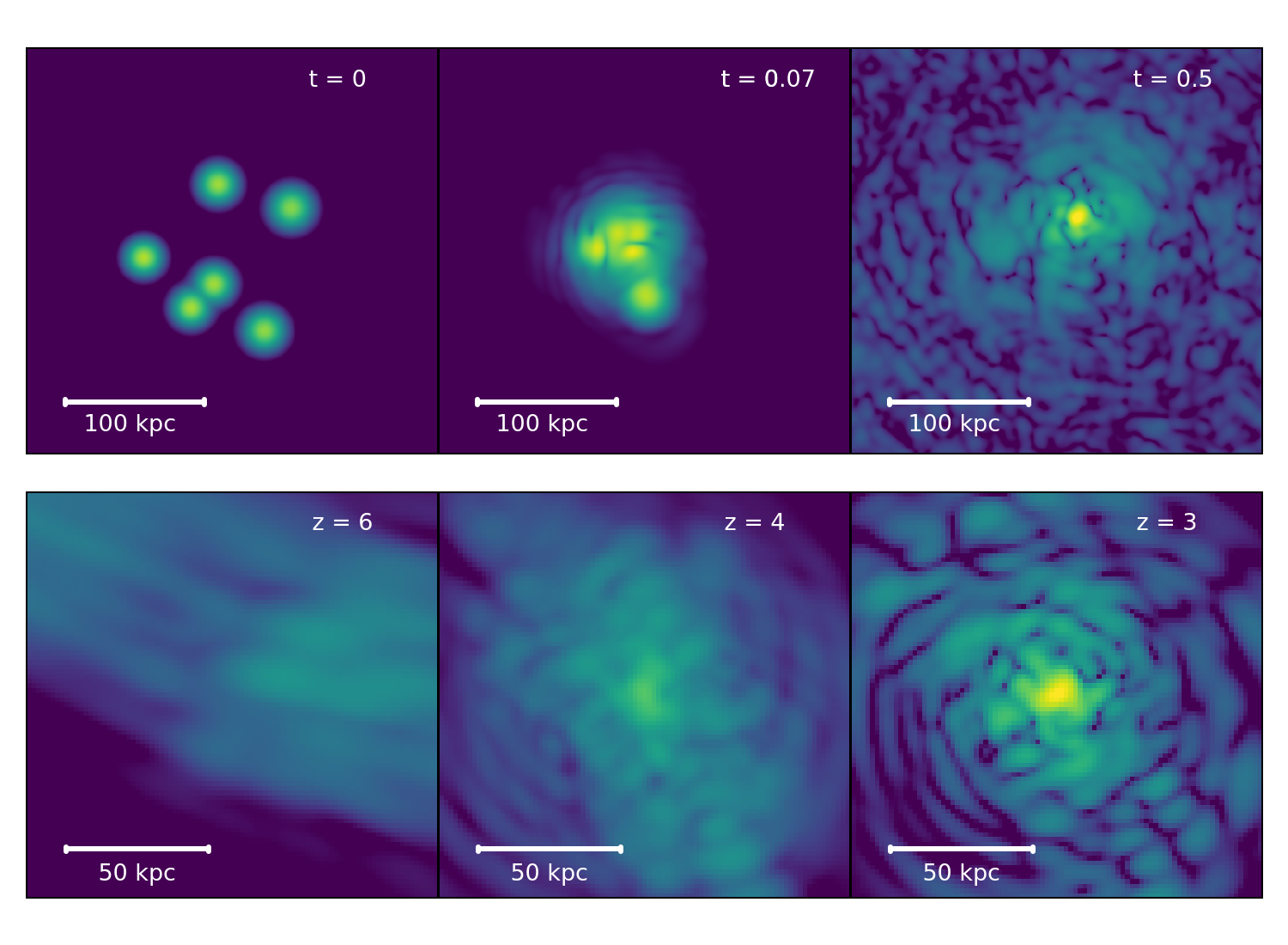}
    \caption{Time evolution of core and halo. The top row shows an example of a soliton merger simulation at $z = 3$ in a box of size \SI{300}{\kpc} with particle mass $mc^2 = \SI{e-22}{\eV}$. The bottom row shows a selected halo formation from the large-scale structure simulations by \citet{Simon2021}. A stable core–halo structure can always be found at the end of all simulations. For illustrative purposes, the first two columns show the projected density (obtained by integrating density slices along the $z$-axis), but the last column is a single slice (i.\,e.\ one grid cell in thickness) of the snapshot through the $z$-coordinate of the halo center.}
    \label{fig:SM}
\end{figure}

\subsubsection{Soliton merger simulations}
The soliton merger simulations are performed with a particle mass $mc^2 = \SI{e-22}{\eV}$, a box size $L = \SI{300}{\kpc}$ and at $z = 3$ on a grid with $N^3 = 512^3$ cells.
The simulations are started with six randomly-placed solitons with mergers mostly occurring at $t \sim 0.1 \, t_\mathrm{H}$, where $t_\mathrm{H}$ is the Hubble time.
Since the simulations at $z = 3$ take 16 times longer than those at $z = 0$ due to the dependence of time step on the scale factor as shown in \cref{eq:timestep}, we stop the simulations at $0.5 \, t_\mathrm{H}$.
We have checked that haloes at $t \sim 0.5 \, t_\mathrm{H}$ are relaxed, since they meet the virialisation criterion $|2(K+Q)/W| \approx 1$ \citep{Hui:2016ltb,Mocz2017} (where $K$, $Q$ and $W$, are the kinetic, quantum and potential energies, respectively).
However, we also included unrelaxed haloes in between $0.1 \, t_\mathrm{H} < t < 0.5 \, t_\mathrm{H}$ in our results.
Alternative initial settings were tested, such as increasing the number of solitons with a larger range of masses, but the results do not change the main conclusion of this work.

\subsubsection{Small-volume cosmological simulations}
A series of cosmological simulations are performed using the same resolution, particle mass and box size.
They all begin from $z = 50$ and stop at $z = 0$.
The initial conditions are generated using \textcode{MUSIC} \citep{MUSIC} with the \ac{CDM} transfer function from \citet{Eisenstein_Hu97,Eisenstein_Hu99}, and the following cosmological parameters: $\Omega_m = 0.276$, $\Omega_\Lambda = 0.724$, $h = 0.677$ and $\sigma_8 = 0.8$.
Due to the difficulty of simultaneously resolving the large-scale structure and the inner non-linear evolution of haloes on a grid size of $512^3$, we use initial conditions that correspond to \q{zoom-in} regions with $L = \SI{300}{\kpc}$ of a larger \SI{1}{\Mpc} box generated by \textcode{MUSIC} with different random seeds.

\subsubsection{Large-volume cosmological simulation}
A large-volume high-resolution cosmological simulation was performed by \citet{Simon2021} with similar cosmological parameters, but larger box size $L = \SI{10}{\Mpc\per\hHubble}$ and grid size $N^3 = 8640^3$, and slightly lighter particle mass $mc^2 = \SI{7e-23}{\eV}$.
With such a box size and spatial resolution, this simulation contains a population of haloes with diverse formation histories, including tidally stripped, isolated, and merged haloes.
Therefore, it provided us with a more realistic measurement of the core–halo mass relation in a \ac{FDM} universe.
\Cref{fig:SM} visually shows the time evolution of the density distribution in different simulations.
It is clear that, whether a halo is formed through soliton mergers or gravitational collapse of large-scale structure, there always exists a stable core structure enveloped by interference fluctuations within its host halo, but we will see later that different box sizes can lead to different types of core–halo structure.

\subsubsection{Initial power spectrum}
As noted above, in this work \citep[as well as][]{Simon2021}, we did not use the initial power spectrum of the \ac{FDM} model, which presents a suppression of power on small scales, because the inner structure of haloes should be insensitive to the initial conditions \citep{Schive2014b}. 
Although different merger histories may lead to different core–halo structure, the extent of this impact is still to be determined.
We assume here that the increased amount of small-scale structure, as well as the number of system interactions, will have negligible effects on the statistics of core–halo structure.
Simulated haloes with comparable size of the soliton are rare if a more realistic power spectrum is applied, but should still exist and therefore be included in the resulting core–halo mass relation.

\subsection{Spatial resolution}

\begin{figure}
	\includegraphics[width=\columnwidth]{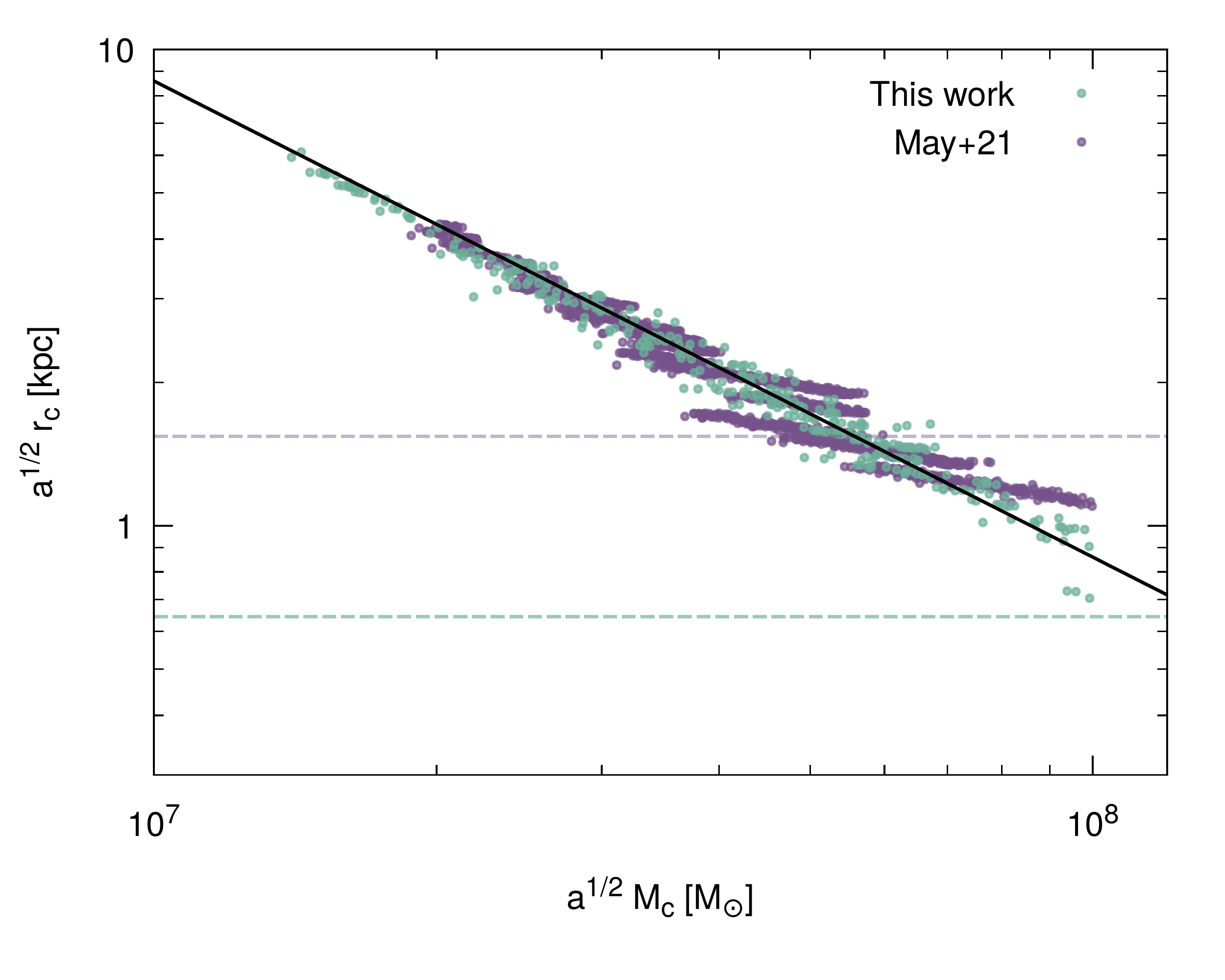}
    \caption{Core mass–radius relation scaled to $mc^2 = \SI{8e-23}{\eV}$ via \cref{eq:scaling}. The black line is a fitting relation \eqref{eq:Mcrc} from \citet{Schive2014a}. The dashed lines show $2\Delta x$ as a reference of the resolution limit for the simulations of this work and \citet{Simon2021}.}
    \label{fig:figMcrc}
\end{figure}

Our soliton merger simulations have a smaller box size, but the same number of grid cells ($512^3$) as our cosmological simulations, so the resolution $\Delta x = \SI{0.644}{\kpc}$ is better than previous studies \citep{Bodo2016,Mocz2017}.
This allows us to resolve smaller cores, but the haloes may experience stripping effects from their own gravitational pull.
On the other hand, although the large simulation is performed in high resolution, the (re-scaled) grid resolution $\Delta x = \SI{1.547}{\kpc}$ is still twice as large as that of the soliton merger simulations.
The importance of resolving the core with fine enough grids is reflected in the core mass–radius relation.
\Cref{fig:figMcrc} shows that simulated haloes have cores following a tight relation:
\begin{equation}
    a^{1/2} M_\mathrm{c} = \frac{\num{5.5e9}}{(mc^2/\SI{e-23}{\eV})^2(a^{1/2} r_\mathrm{c}/\si{\kpc})} \Ms\,.
    \label{eq:Mcrc}
\end{equation}

As the core becomes more massive, the core size decreases further.
When the core size is resolved by less than two grid cell lengths, the relation becomes more dispersed and discretised.

\section{Results}
\label{sec:results}

\subsection{Density profiles}

\begin{figure}
	\includegraphics[width=\columnwidth]{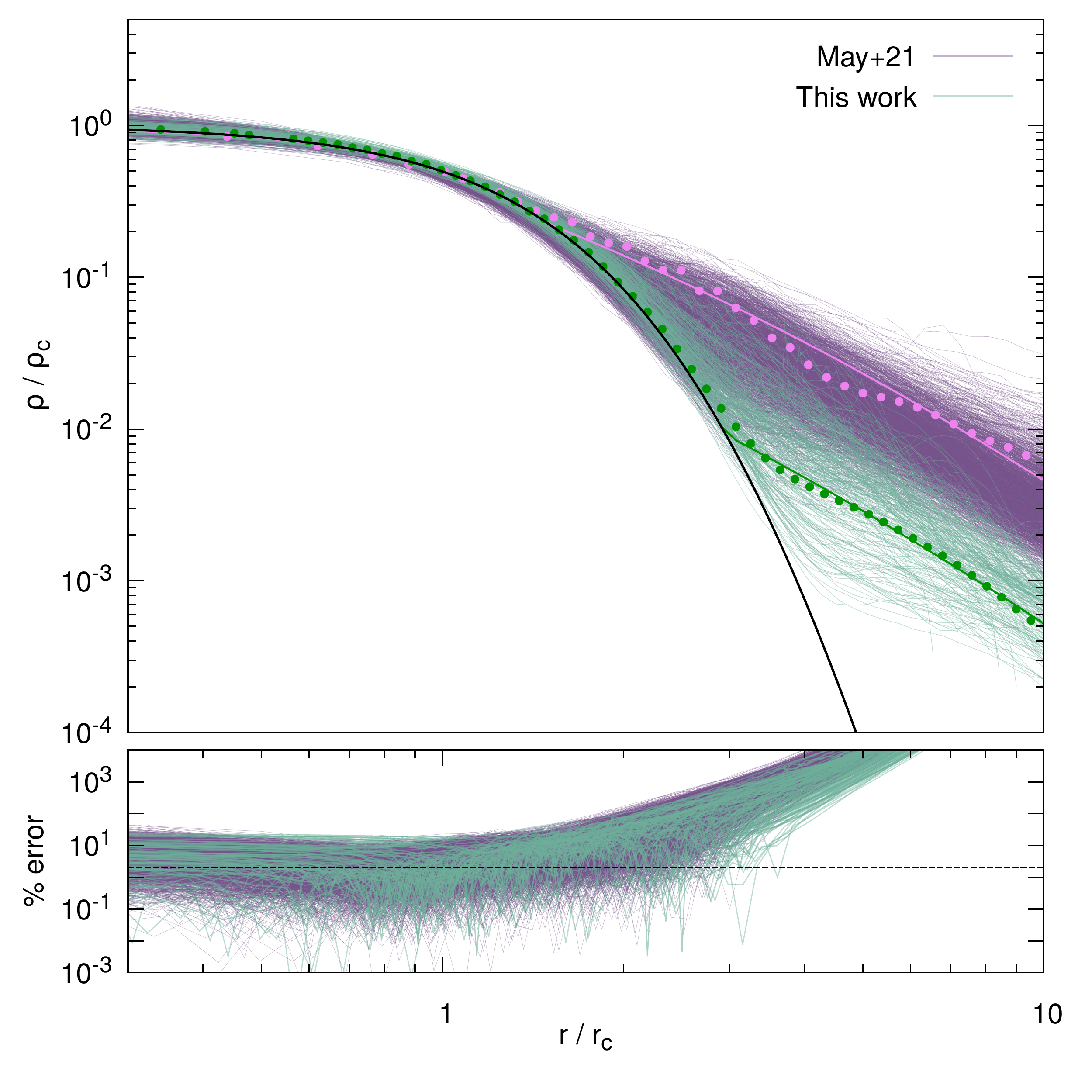}
    \caption{Scaled density profile of haloes obtained from simulations of this work and \citet{Simon2021}. The scaled core profile is shown as black line. We highlight two haloes with pink and dark green and their best-fit cored \ac{NFW} profile. They have similar core mass, but an order of magnitude difference in the halo mass. Bottom sub-panel shows the percentage error between data and core profile. The dashed line denotes an error of \SI{2}{\percent}.}
    \label{fig:profiles}
\end{figure}

The centres of the haloes from the simulations performed in this work are found by the minimum gravitational potential, and those from the cosmological simulation in \citet{Simon2021} are determined by selecting the densest cells of haloes found by a grid-based friends-of-friends-like halo finder.
We measured the spherically averaged density profile and performed fitting to \cref{eq:corenfw} to extract $r_\mathrm{c}$, $r_\mathrm{t}$ and $r_\mathrm{s}$ for all haloes.
As shown in \cref{fig:profiles}, a flat cored structure is identified towards the center in all profiles.
They are well fitted by the core density profile \cref{eq:corenfw} with a maximum error of \SI{10}{\percent} up to the core radius $r_\mathrm{c}$.
After the transition radius $r_\mathrm{t}$, the profiles follow the \ac{NFW} profile.
We also see that for some haloes, we have a direct transition from the core to the \ac{NFW} profile, while others show a longer transition with an intermediate behaviour linking the two regimes.

One interesting feature we observe is oscillations in these profiles in their outer regions that can only be modelled on average by the smooth \ac{NFW} profile.
A possible reason for the fluctuations is that they are caused by the interference granules in the \ac{NFW} region.
If this is true, it is possible that halo density profiles can be used to measure this unique interference pattern present in models like \ac{FDM}.
More tests are needed to confirm this hypothesis.

\begin{figure}
	\includegraphics[width=\columnwidth]{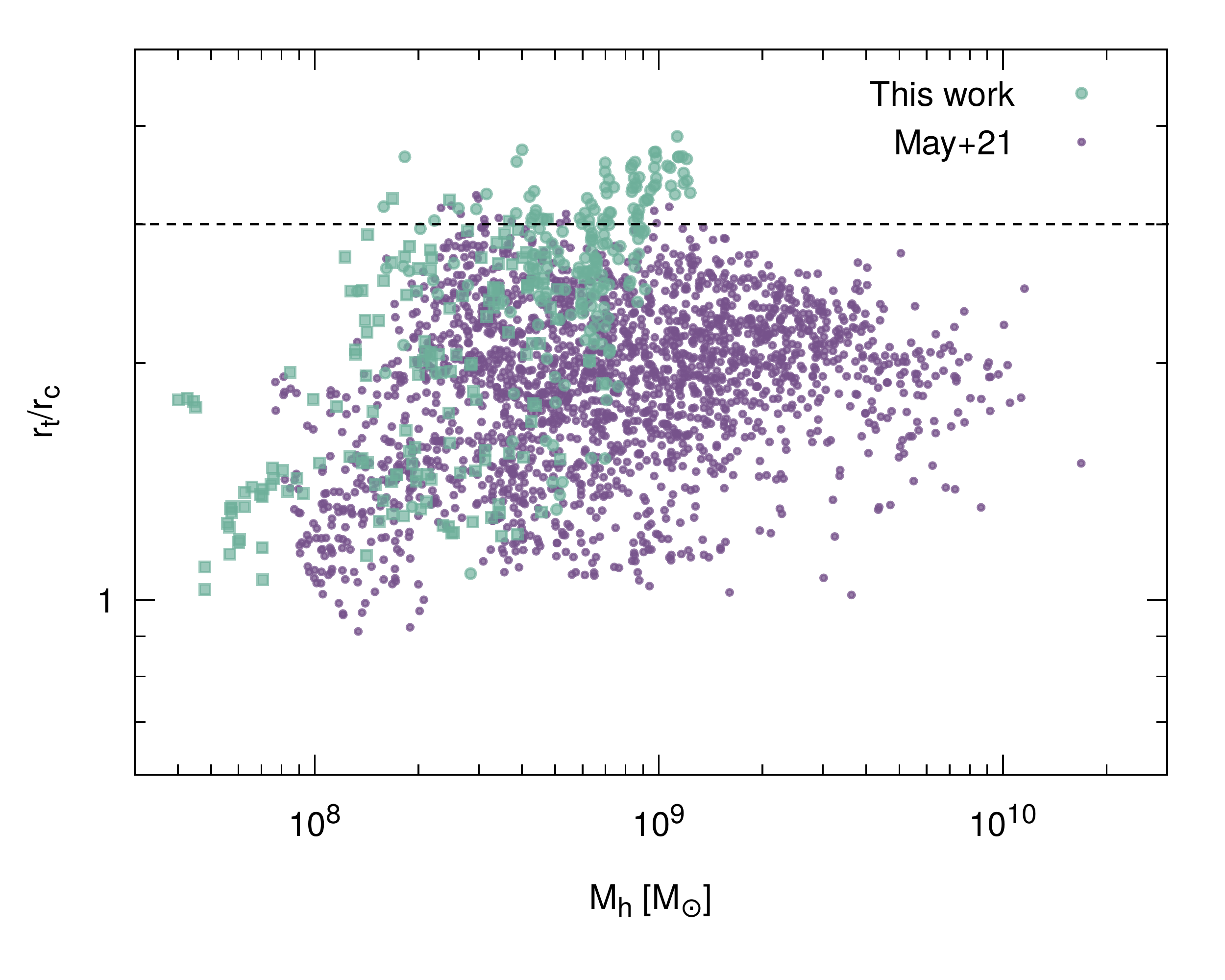}
    \caption{Range of transition radius as a function of halo mass. The dashed line shows the typical transition $r_\mathrm{t} = 3r_\mathrm{c}$ obtained from \citet{Schive2014b}.}
    \label{fig:rtMh}
\end{figure}

In previous simulations \citep{Schive2014b,Mocz2017}, the transition radius was found to be $r_\mathrm{t} \ge 3r_\mathrm{c}$, where the residual error between the data and the core profile is less than \SI{2}{\percent} for $r < r_\mathrm{t}$.
However, our measured $r_\mathrm{t}$, purely from fitting to the cored \ac{NFW} profile \cref{eq:corenfw}, disagrees with these previous results.
The error at $3r_\mathrm{c}$ is greater than at least \SI{10}{\percent}, as shown in the bottom panel of \cref{fig:profiles}, meaning the actual $r_\mathrm{t}$ should be located at a radius smaller than $3r_\mathrm{c}$.
The range of values for the measured $r_\mathrm{t}$ in \cref{fig:rtMh} shows that most haloes do have $r_\mathrm{t} \le 3r_\mathrm{c}$.
Other recent work, such as \citet{Yavetz2021}, also shows smaller transition radii, e.\,g.\ $r_\mathrm{t} \approx 2r_\mathrm{c}$.
As mentioned before, from theory, to guarantee a continuous and smooth transition from the solitonic core to the \ac{NFW} profile, continuity of both the density and of its first derivative would be necessary, which translates to the requirement $r_\mathrm{t} \le 3r_\mathrm{c}$, which, therefore, agrees with our result.
This implies that all the haloes in the simulations presented here have a continuous and smooth transition from the core to the \ac{NFW} profile, with or without a transition period, and thus do not suffer from the apparent inconsistency present in previous simulations.

\begin{figure}

    \includegraphics[width=\columnwidth]{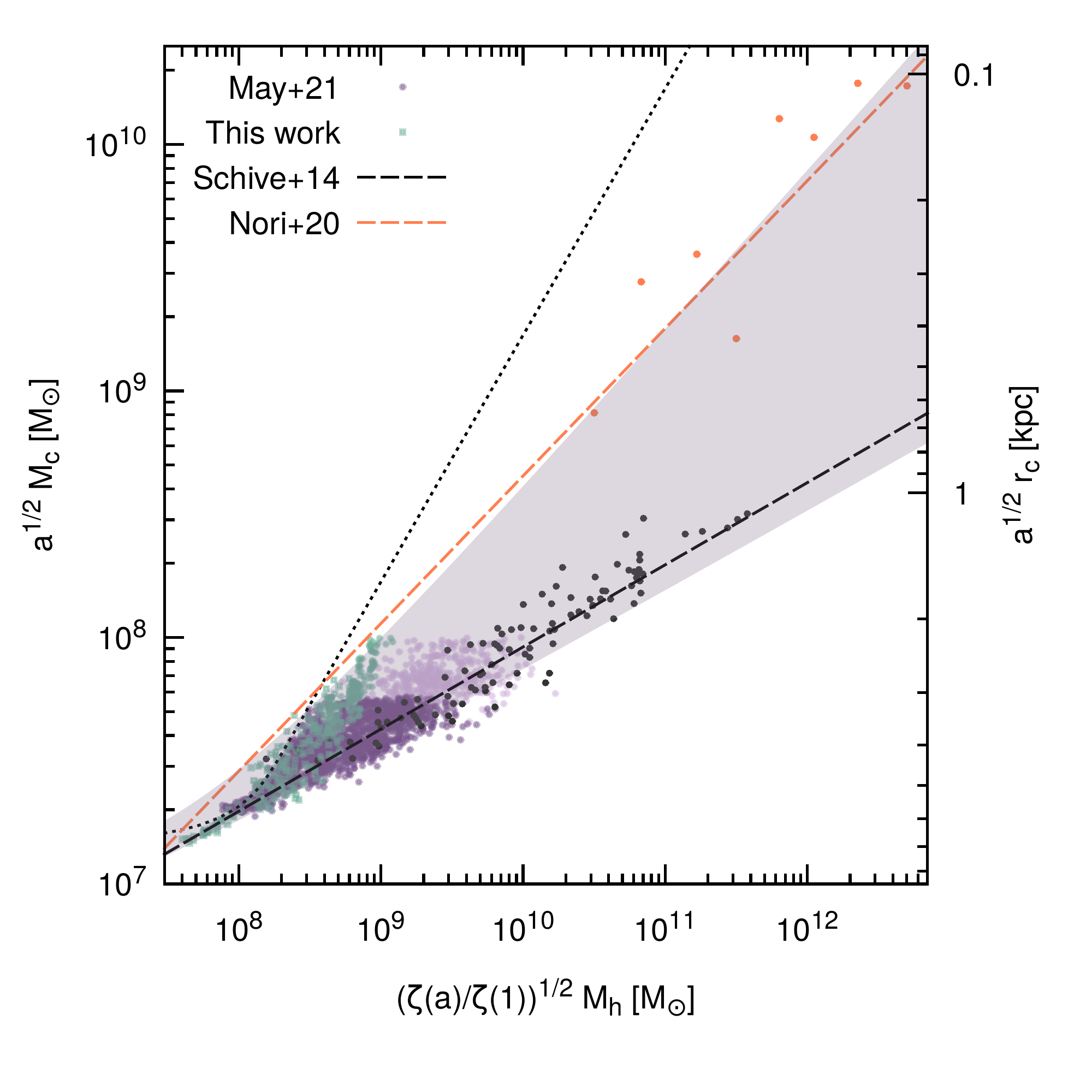}
    
    \caption[We adopted parameters resulting from the varying exponents analysis without sub-sampling restrictions by dynamical and morphological information.]{Core–halo relation scaled to $mc^2 = \SI{8e-23}{\eV}$ via \cref{eq:scaling}. Green dots are haloes simulated in this work with cores resolved by at least $3\Delta x$. Purple and and faint purple dots are haloes from the large-box cosmological simulation \citep{Simon2021} with cores resolved by at least 2$\Delta x$ and $\Delta x$ respectively. The pink shaded region is enclosed by the empirical fits to the purple and green dots, with the maximum and minimum values of the parameters in \cref{eq:chrelation}. The solid dotted line corresponds to the soliton-only relation obtained from a pure core profile. The black and orange dashed lines are fitting relations corresponding to the black and orange dots obtained from \citet{Schive2014b} and \citet{Nori2021}\footnotemark respectively.}
    \label{fig:chrelation}
\end{figure}

\subsection{The core–halo mass relation}
\label{sec:results-core-halo-relation}

\Cref{fig:chrelation} shows the core–halo mass relation obtained from the soliton merger and cosmological simulations.
All data are scaled to $mc^2 = \SI{8e-23}{\eV}$ using \cref{eq:scaling} in order to enable a direct comparison with the data and fitting relation from \citet{Schive2014b}.
For reference, we also show the \q{core–halo} mass relation of a soliton-only profile, i.\,e.\ a pure core profile with $r_{\mathrm{t}} \to \infty$ in \cref{eq:corenfw}, represented by the solid black line.
This curve indicates the minimum halo mass for a certain core mass, and any haloes located to the right of the soliton-only core–halo relation must have the usual cored \ac{NFW} structure.
For haloes in the soliton merger simulations with mass $\gtrsim \SI{e8}{\Ms}$, the relation has a steeper slope than $\alpha = 1/3$, confirming the results from \citet{Mocz2017}.
However, haloes from the large-scale cosmological simulation predict a core–halo relation with a large enough dispersion that can cover a range of data produced by both the soliton merger simulations and \citet{Schive2014b}.
The range of the dispersion can span as large as one order of magnitude in halo mass for $M_\mathrm{c} \sim \SI{5e7}{\Ms}$.
This dispersion, which fills in the space in between the soliton-only line and the relation from \citet{Schive2014b}, indicates the diversity of the cored \ac{NFW} structure in the \ac{FDM} simulations.
For example, \cref{fig:profiles} highlights two profiles of haloes with similar core mass $M_\mathrm{c} \sim \SI{5e7}{\Ms}$, but different halo mass.
The tight \q{one-to-one} core–halo relations found by different groups, with different slopes, therefore only describe a part, but not all populations of haloes in the \ac{FDM} model.

We suggest an empirical equation that has the following form: $M_\mathrm{c} = \beta + \left( M_h/\gamma \right)^\alpha$.
The parameter $\beta$ takes the limit of the relation for small halo masses into account, although low-mass haloes are rare in a \ac{FDM} universe due to the suppression in the initial power spectrum.
$\alpha$ is the slope that can be compared to previous works.
After including the scaling symmetry in \cref{eq:scaling} and the redshift dependence according to \citet{Schive2014b}, we have
\begin{align}
\begin{split}
    a^{1/2} M_\mathrm{c} &= \beta \left(\frac{mc^2}{\SI{8e-23}{\eV}}
    \right)^{-3/2}
    \\
    &+ \left( \sqrt{\frac{\zeta(z)}{\zeta(0)}}\frac{M_\mathrm{h}}{\gamma}\right)^\alpha \left(\frac{mc^2}{\SI{8e-23}{\eV}}
    \right)^{3(\alpha-1)/2} \Ms\,.
    \label{eq:fitJ}
\end{split}
\end{align}
The best-fit parameters for the haloes from the large-box cosmological simulation give
$\beta = 8.00^{+0.52}_{-6.00} \times 10^{6} \, {\Ms}, \log_{10}(\gamma / {\Ms}) = -5.73^{+2.38}_{-8.38} \, $ and $\alpha = 0.515^{+0.130}_{-0.189}$,
which is shown as a pink shaded region in \cref{fig:chrelation}.

\footnotetext{We adopted parameters resulting from the varying exponents analysis without sub-sampling restrictions.}

The effect of the large dispersion is encompassed in the uncertainty of the model parameters.
This uncertainty is not the statistical uncertainty of the fit, but an \q{overestimation} of the uncertainty in the parameters that can reflect the large dispersion of the data.
Indeed, the statistical uncertainty would be the incorrect quantity to consider in this case, since we do not assume that there is an underlying \q{true} set of values for the parameters with statistical fluctuations, but rather propose that different halo populations could \emph{systematically} follow different relations depending on their histories and properties (see \cref{sec:origin-disperion}).
To obtain a more appropriate description of the core–halo diversity, we employed \ac{KDE}, estimating the probability distribution function of the core masses with respect to the central value of the corresponding binned halo mass.
Each of these distributions reveals the dispersion of core masses for each halo mass.\footnote{%
    We can provide the distribution of core masses for each halo mass bin by request for those interested.%
}
We then obtain the minimum and maximum curves $M_{\mathrm{c}}(M_{\mathrm{h}})$ that fit all of these distributions, and extract the minimum and maximum vales for the parameters $b$, $\gamma$ and $\alpha$ from these curves.
The difference to the global fit is our uncertainty in the parameters.

\citet{Nori2021,Mocz2017,Schive2014b} determined slopes $\alpha$ of \numlist{0.6; 0.556; 0.333}, respectively.
Given the large dispersion seen in our data, all of these slopes are compatible when taking into account the uncertainty we assigned to the fitting parameters.
So when considering the fitting function we propose, all of the other cases in the literature are covered as well.
We emphasise that our results show that a general halo population is not well-described by any single one-to-one core–halo mass relation.
Further investigation is required to determine which halo populations follow which relations (if any), and under what conditions – cf.\ \cref{sec:origin-disperion}.

This large spread and uncertainty in the fitting function can affect the constraints on the \ac{FDM} mass obtained from these relations.
Here, we provide a rough estimate of the error.
For the same halo mass $M_\mathrm{h} = \SI{e9}{\Ms}$ in \cref{fig:chrelation}, we can have the least massive core mass as $M_\mathrm{c} = \SI{3e7}{\Ms}$ and the most massive as $M_\mathrm{c} = \SI{e8}{\Ms}$. Applying these values to the core density in \cref{eq:corenfw} gives a \SI{50}{\percent} difference in particle mass $m$.
Therefore, any observational constraints made using the relation \cref{eq:chrelation} should include an additional uncertainty on the order of \SI{50}{\percent} in the results, unless the halo mass is smaller than $10^9 (\SI{8e-23}{\eV}/(mc^2))^{3/2} \,\Ms$.
Therefore, when obtaining the \ac{FDM} mass using the core–halo relation, one needs to take into account the dispersion of these values, shown in the uncertainty in the fitting parameters, which will translate to a higher uncertainty in the \ac{FDM} mass.

We now scrutinize whether the scatter of the core–halo relation has an influence on the \ac{FDM} mass constraints through a dynamical analysis for dwarf galaxies, as has been performed in the literature when fitting the presence of a core in such galaxies.
To this end, we apply the spherical Jeans analysis to the kinematic data of the Fornax dwarf spheroidal galaxy, which has the largest data set among the Galactic dwarf satellites.
We perform the Jeans analysis\footnote{For the dynamical analysis we adopt in this work, the interested reader may find further details in \citet{Kohei2021}.} using two different core–halo relations, which are suggested by \citet{Schive2014b} and this work, and then we map the posterior probability distributions of the \ac{FDM} mass through the \ac{MCMC} technique based on Bayesian statistics. 
Comparing the posteriors, there is no clear difference in the shape of those distributions, including that of \ac{FDM} mass, but this is due to the fact that there exists a degeneracy between halo mass and \ac{FDM} mass.Therefore, this degeneracy makes it hard to see the impact that the core–halo relation has in the Jeans analysis.

Due to limited spatial resolution, we could only observe the dispersion to increase with halo mass until $M_\mathrm{c} \sim \SI{6e7}{\Ms}$.
It would be important for potential future higher-resolution simulations to examine if the dispersion keeps increasing along the soliton-only relation or not.
Again, the increasing dispersion is of importance to observational studies since it will also lead to an increasing uncertainty in the core–halo relation.

\subsubsection{The origin of the dispersion}
\label{sec:origin-disperion}


Different core–halo structures have been found in different simulations:
\begin{itemize}
    \item As mentioned before, \citet{Schive2014b} and \citet{Mocz2017} find different results for the slope $\alpha$ ($1/3$ vs.\ $5/9$), even for similar simulation setups (soliton mergers).
    \item \citet{Mina2020} claim to confirm a slope of $\alpha = 5/9$, as found in the soliton merger simulations of \citet{Mocz2017}, but using a \emph{cosmological} simulation, contradicting the result of $\alpha = 1/3$ from \citet{Schive2014b}.
    However, the number of haloes in their sample is very small.
    \item \citet{Bodo2016} performed soliton merger simulations similar to \citet{Schive2014b} \citep[and later][]{Mocz2017}\footnote{%
        Although \citet{Bodo2016} made use of \q{sponge} boundary conditions instead of periodic boundary conditions.%
    } and could not reproduce the previously-found value of the slope $\alpha$, or indeed any universal relation.
    \item \citet{Nori2021} studied the dynamics of eight simulated haloes and concluded with a similar comment: \citet{Schive2014b} and \citet{Mocz2017} only captured a partial representation of the core–halo relation in a realistic cosmological sample.
    \item \citet{Yavetz2021} used the Schwarzschild method to construct self-consistent \ac{FDM} halos and found that a stable core–halo structure can exist even when the adopted core–halo mass relation deviates from \citet{Schive2014b}.
\end{itemize}
These examples illustrate that the diversity of the possible core–halo slopes found in different works seems originate from the type of simulations performed, which results in halos and cores that have different properties.
The diversity of core–halo structure found in these simulations is exhibited in our work, where we can clearly see the difference between the core–halo mass relation from halos formed in soliton merger simulations (green points in \cref{fig:chrelation}) and in cosmological simulations (pink points in \cref{fig:chrelation}).

We can think of a few possible explanations for this diversity of halos: merger history \citep{Du:2016aik,Yavetz2021}, tidal stripping effects, and the relaxation state of the halo \citep{Nori2021}. 
Formation and merger history is an explanation that seems very plausible to be a relevant factor.
Larger cosmological simulations, like the one from \cite{Simon2021}, present halos that could have very different merger histories, and a large dispersion is expected.
This is different from the soliton merger simulations, where we do not expect a complicated merger history.
We leave for future work to try to identify the different merger histories and try to clarify how this relates to the different incarnations of the core–halo mass relation.


Another possible factor that can also contribute to the dispersion found is stripping.
Here, we will attempt to provide an argument to support tidal stripping as one element responsible for the dispersion, based on the setups of various simulations.
By comparing the box sizes and the resulting slopes $\alpha$ between the small-volume cosmological simulations of this work with \citet{Mocz2017} and \citet{Schive2014b}, which are \SIlist{335; 1765; \ge 2000}{\kpc} (box sizes) after re-scaling via \cref{eq:scaling}, and \numlist{\sim 0.9; 0.556; 0.333} (slopes) respectively, we find that smaller simulation box sizes are correlated with a steeper slope in the core–halo relation.
This can be explained by the stripping effect on the halo by its own gravity due to the periodic boundary conditions:
the self-stripping effect becomes more effective at removing mass from the \ac{NFW} region as the box size decreases.
This skews the core–halo structure towards smaller halo masses, steepening the core–halo relation.
A more rigorous test to prove the above argument requires simulations with increased spatial resolution and box sizes up to at least \SI{2}{\Mpc}, which current numerical schemes are unable to feasibly achieve.

The self-stripping effect is a numerical artifact, but there is no doubt that a stable core–halo structure can exist within such environments.
In more realistic cosmological simulations, dwarf satellites also experience a similar effect from their host haloes in the form of tidal stripping.
Therefore, we suggest that stripping effects by tidal forces are one of the contributing factors causing the dispersion obtained from the large-box simulation in \citet{Simon2021}.
One subtlety is that the tidal effect is an interaction between host haloes and sub-haloes with at least two orders of magnitude difference in mass, but the halo finder used in \citet{Simon2021} does not identify sub-haloes.
However, it is known that sub-haloes in \ac{CDM} simulations can temporarily move outside of the virial radius of the host halo after the first pericentric passage \citep{Bosch2017}.
We assume that ejected sub-haloes should also exist in a \ac{FDM} cosmology, and therefore identified by the halo finder.
An in-depth analysis of the tidal effect on the core–halo relation, or \ac{FDM} sub-haloes in general, would require building merger trees, which is still not yet studied in any \ac{FDM} cosmological simulations. We leave this investigation to future work.

\subsection{Other relations}
\subsubsection{Inner dark matter slope–halo mass relation}

Observational constraints obtained through Jeans analysis require adopting the cored \ac{NFW} density profile and core–halo mass relation.
The scatter in the core–halo mass relation plays a part in the analysis simply as an uncertainty in the relation.
To study the observational consequences of the diversity, we suggest showing the inner slope–halo mass relation and core radius–halo mass relation for our \ac{FDM} haloes, which can be compared to previous observational results. 

\begin{figure}
	\includegraphics[width=\columnwidth]{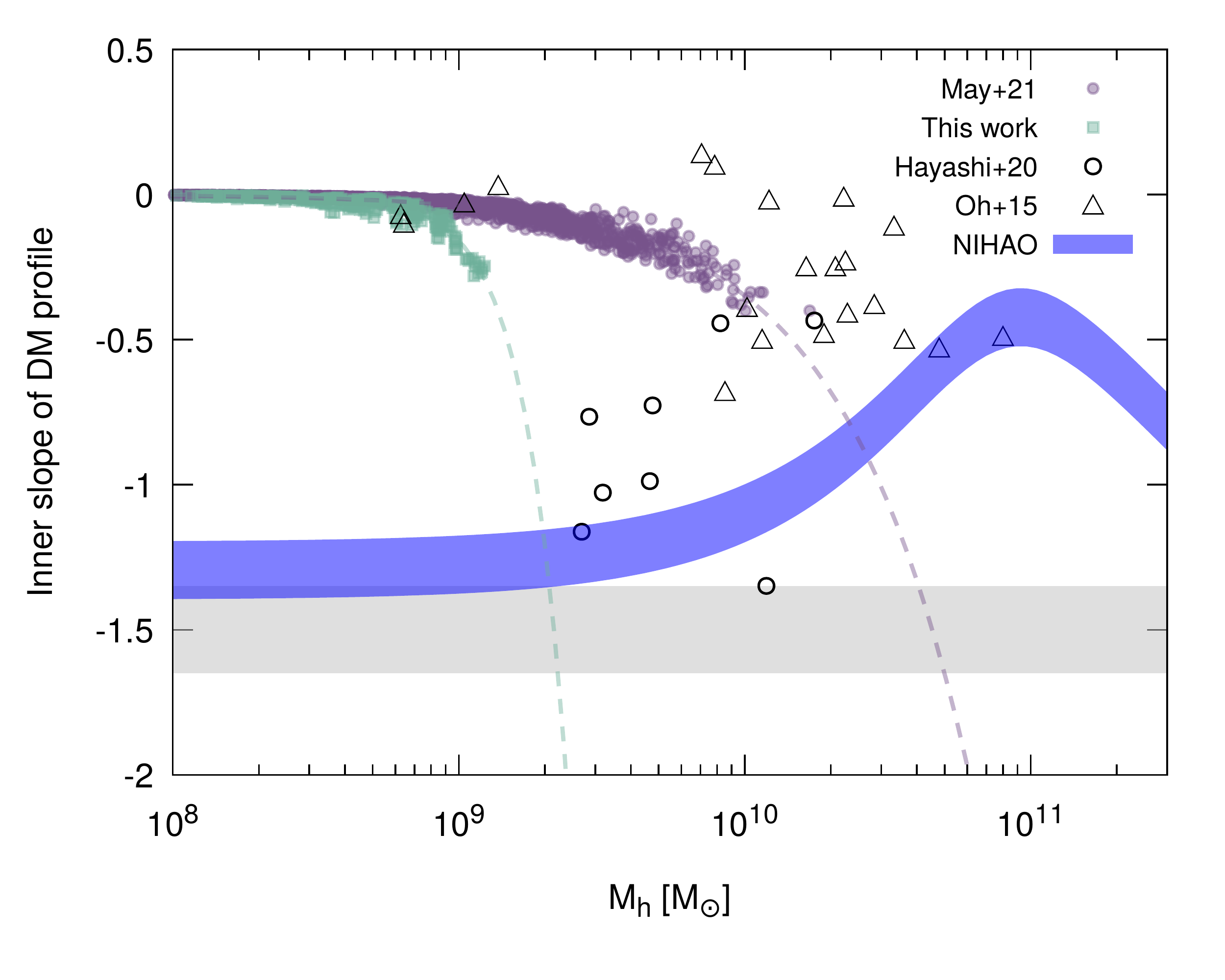}
    \caption{Inner dark matter slope as a function of halo mass. The inner slope is defined as the logarithmic gradient density at $\num{0.015} r_\mathrm{h}$. Green and purple dots represent haloes from simulations of this work and \citet{Simon2021}, where halo mass is rescaled to $mc^2 = \SI{8e-23}{\eV}$ and $z = 0$ via \cref{eq:scaling}. Open triangles are the observed relation from dwarf galaxies based on Jeans analysis \citep{Kohei2020}, whereas open circles are predicted from the rotation curves of dwarf galaxies \citep{Oh2015}. The blue band is a fitting function with an uncertainty of $\pm 0.1$ predicted by NIHAO, a \ac{CDM} simulation with baryonic feedback physics \citep{Tollet2016}. The grey band shows the prediction by \ac{CDM}-only simulations.}
    \label{fig:slopeMh}
\end{figure}

We define the inner slope as the logarithmic gradient dark matter density $\Delta\log \rho/\Delta\log r$ at an inner radius of \SI{1.5}{\percent} of the halo's virial radius $r_\mathrm{h}$: $r_\text{inner} = \num{0.015} r_\mathrm{h}$.
The definition is frequently used to study the impact of feedback physics on the inner dark matter structure \citep{Tollet2016}.
As shown in \cref{fig:slopeMh}, the inner slope of \ac{FDM} haloes is expected to be cored (i.\,e.\ 0) for less massive haloes with mass $\lesssim \SI{e9}{\Ms}$.
In contrast, haloes in \ac{CDM} simulations with baryonic feedback physics show a cuspy inner slope $\sim -1.5$ within this mass range, due to the inefficient core formation process by feedback \citep{Tollet2016}.
It is therefore important to observe the inner slope of ultra-faint dwarf galaxies, which can help to distinguish between feedback-induced and quantum pressure-induced cores.
As the relation moves to more massive \ac{FDM} haloes, the inner radius begins to shift outside of the cored region because of the inverse proportionality between core radius and halo mass.
As a result, the inner slope steepens.
Note that the steepening occurs at different halo mass ranges for different sets of \ac{FDM} halo samples, because haloes in simulations of smaller box size tend to be stripped, so the steepening occurs earlier.

The inferred observational relation from the stellar kinematics of eight dwarf galaxies \citep{Kohei2020}, and rotation curves of 26 dwarf galaxies \citep{Oh2015} shows a large scatter of inner slope for a certain halo masses, which is a result of diverse dark matter density profiles.
If we consider an extrapolation of the inner slope–halo mass relation (dashed lines in \cref{fig:slopeMh}), including both small and large box size simulations, the \ac{FDM} model with $mc^2 \approx \SI{8e-23}{\eV}$ may be able to explain the scatter presented by the observations.

We caution that the definition of halo mass and inner slope vary across the literature.\footnote{%
    As another detail, the halo masses of the purple and green points in \cref{fig:slopeMh} are extracted at $z = 3$ and rescaled to $z = 0$ with the factor $(\zeta(z=3)/\zeta(z=0))^{1/2}$, which corresponds to a mass of $M_\mathrm{h} = 350 \, \rho_\mathrm{m0}\, (4\pi r_\mathrm{h}/3)$, whereas all other data in \cref{fig:slopeMh} used $M_\mathrm{h} = 200 \,\rho_\mathrm{m0} \,(4\pi r_\mathrm{h}/3)$.
    Changing the definition would simply shift the data horizontally in \cref{fig:slopeMh}.%
}
Moreover, populating the region in between the extrapolated relations would require sub-halo data, which we did not investigate in this work.
We therefore emphasise that the particle mass $mc^2 \approx \SI{8e-23}{\eV}$ only represents a loose constraint, and the main motivation of \cref{fig:slopeMh} is to demonstrate the possibility of explaining the observed diversity of inner slopes by stripped, or more realistically, tidally stripped sub-haloes, which is closely related to the diversity of the core–halo structure.

\subsubsection{Core radius–halo mass relation}

\begin{figure}
	\includegraphics[width=\columnwidth]{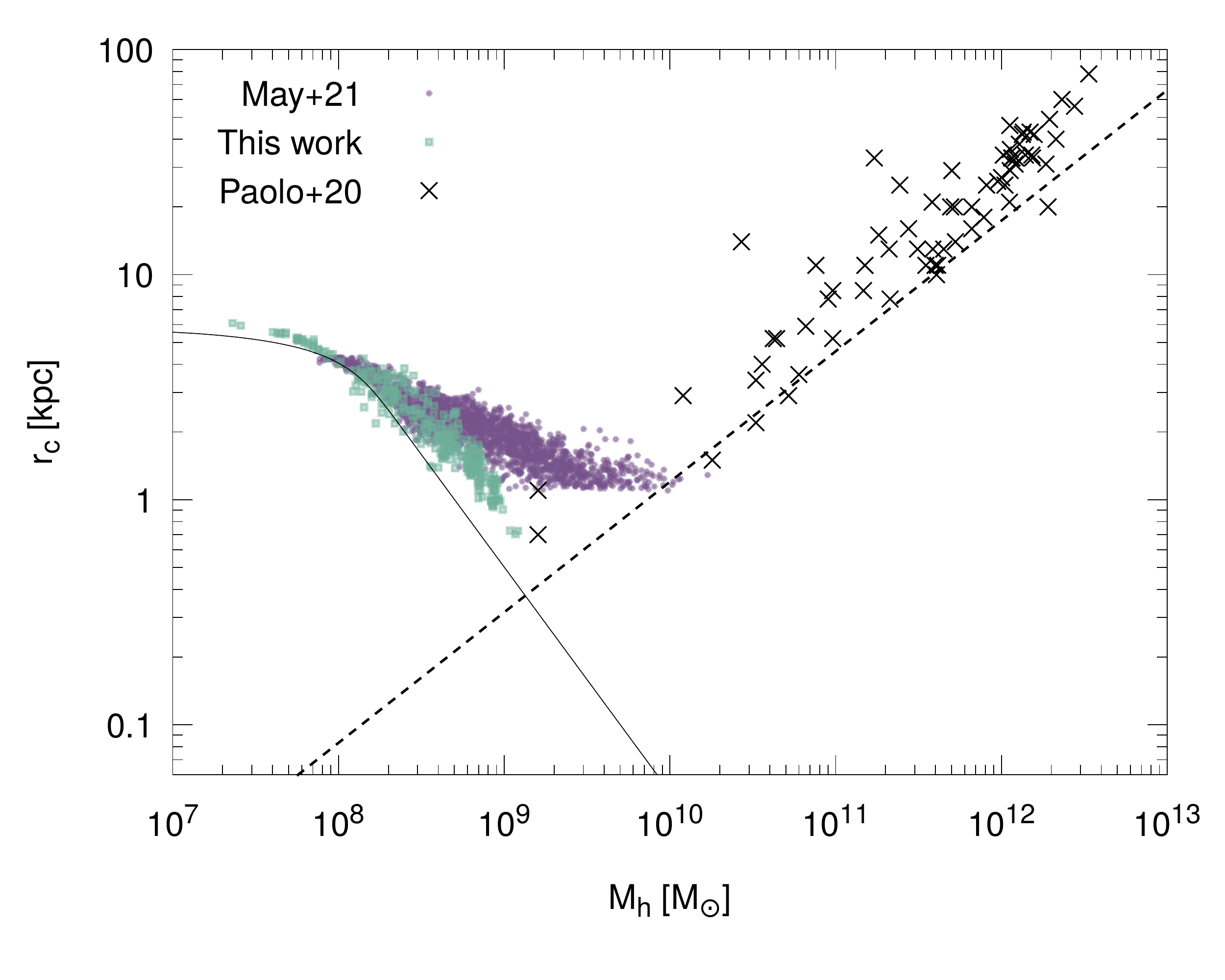}
    \caption{Core radius vs.\ halo mass. Green and purple points are properties of haloes from simulations of this work and \citet{Simon2021}. The black line shows the relation predicted by a soliton-only density profile. The dashed line is an empirical function predicted by \ac{LSB} galaxies \citep{Salucci2007}. Black crosses are from \citet{Paolo2019}.}
    \label{fig:rcMh}
\end{figure}

As suggested by \citet{Burkert2020}, the \ac{FDM} model may fail to explain the observed trend of the core radius–halo mass relation measured from dwarf galaxies.
We follow \citet{Mina2020} and present the core radius–halo mass relation measured from our \ac{FDM} halo samples.
As shown in \cref{fig:rcMh}, the scatter is still observed, but the decreasing trend, which is a fundamental property of quantum pressure-induced cores, is in disagreement with the positive scaling predicted by \ac{LSB} galaxies \citep{Salucci2007,Paolo2019}.

The disagreement is expected because the negative scaling, where less massive galaxies are cored, allows the \ac{FDM} model to solve the core–cusp problem, but the relation from \ac{LSB} galaxies has the opposite behaviour, where massive galaxies have larger cores.
In addition, \ac{LSB} galaxies are predicted in \ac{CDM} simulations to have experienced tidal heating and supernova feedback \citep{Martin2019}.
Therefore, the relation between core radius and halo mass poses a challenge to the \ac{FDM} model, but more importantly, it motivates future \ac{FDM} simulations to include baryonic physics to verify if \ac{LSB}-like galaxies can be formed or not.

\section{Conclusion}
\label{sec:conclusion}

Solitonic cores are found to be formed in simulations of the \ac{FDM} model as a consequence of gravity and the uncertainty principle, but there is still no consensus on a single universal scaling relation that describes the relationship between a halo's mass and that of its core, or that one even exists.
In this work, we performed new soliton merger simulations and used data from a large-scale cosmological \ac{FDM} simulation. 
All simulations are evolved by solving the Schrödinger–Poisson equations through the pseudo-spectral method, which can capture wave phenomena completely.
Here is a summary of our findings.

We found an agreement between the measured density profiles and a cored \ac{NFW} profile, but the transition radii of most of haloes are located at $\le 3r_\mathrm{c}$.
This is in disagreement with previous simulations \citep{Schive2014b,Mocz2017}, but more consistent with the analytical requirement where the transition between the inner core and the outer \ac{NFW} profile must be continuous and smooth.

The resulting core–halo mass relation, obtained from both soliton merger and cosmological simulations, shows an increasing dispersion with halo mass.
The spread extends all the way from the limit of a pure soliton profile to that of \citet{Schive2014b}, signifying the diversity in core–halo structure.
We suggest that, for small cosmological simulations, \q{artificial} stripping effects due to periodic boundary conditions could partially be responsible for the variety of slopes in the relation predicted by different simulations.
However, \q{natural} tidal stripping effects of various severity also exist in larger simulations, which therefore exhibit a greater spread in the relation.
Further, the exact impact of variations between individual haloes on the relation, such as merger history or relaxation state, remains to be uncovered.

We provided a new empirical equation that considers the non-linearity in the low-mass end, but we emphasise that any core–halo relation must suffer from an uncertainty produced by the diversity demonstrated in this work.
Therefore, observational analyses that adopted a core–halo relation must take into account this uncertainty in the fitting parameters, including the particle mass of the \ac{FDM} model.

Due to the limited spatial resolution imposed by the time step criteria, our samples still do not represent the full population of core–halo structure.
To obtain this, simulations using a more flexible numerical scheme, such as adaptive mesh refinement \citep{Schive2014a,Mina2020}, and sub-halo catalogues from merger trees would be needed.
Such future work would provide verification of whether the dispersion keeps growing beyond halo masses of $10^9 (\SI{8e-23}{\eV}/(mc^2))^{3/2} \,\Ms$, or whether the tidally stripped sub-haloes can explain the observed diversity in the inner slope–halo mass relation.
We also plan in the future to understand the merger history of the halos we have in the cosmological simulation, using the same techniques as for \ac{CDM}, in order to try to understand how halos with different merger histories influence the core–halo mass relation.

Lastly, including baryonic physics will further complicate the core–halo structure because the core can now not only be induced by quantum pressure, but also by stellar feedback physics, not to mention the question of how these processes would interact.
However, only baryonic physics have a chance of matching the core radius–halo mass relation of \ac{LSB} galaxies with \ac{FDM}.

\section*{Acknowledgements}
The authors would like to thank Hsi-Yu Schive for providing the data from his simulations.
We would also like to thank Luisa Lucie-Smith for fruitful discussions on the data analysis.
This work was supported in part by MEXT/JSPS KAKENHI Grant Numbers JP21H05448 (for M.C.), 
JP21K13909 and JP21H05447 (for K.H.).
Numerical computations were (in part) carried out on Cray XC50 at the Center for Computational Astrophysics, National Astronomical Observatory of Japan.
The Kavli IPMU is supported by World Premier International Research Center Initiative (WPI), MEXT, Japan.


\section*{Data availability statement}
The data underlying this article will be shared on request to the corresponding authors.


\renewcommand*{\refname}{REFERENCES}
\bibliographystyle{mnras}
\bibliography{FDM_MNRAS}

\newpage
\appendix

\section{Code comparison}
\label{sec:code-comparison}

Since there is no analytical solution to the general time-dependent Schrödinger–Poisson equations, we can only ensure reliability of the code in the general case (beyond toy examples and limiting cases) through comparison with other groups.
Thus, we compared the dark matter density fluctuations at $z = 0$ with \citet{Simon2021} in a test simulation.
Our codes are independently developed, but adopted the same pseudo-spectral splitting method in second order.

\begin{figure}
	\includegraphics[width=\columnwidth]{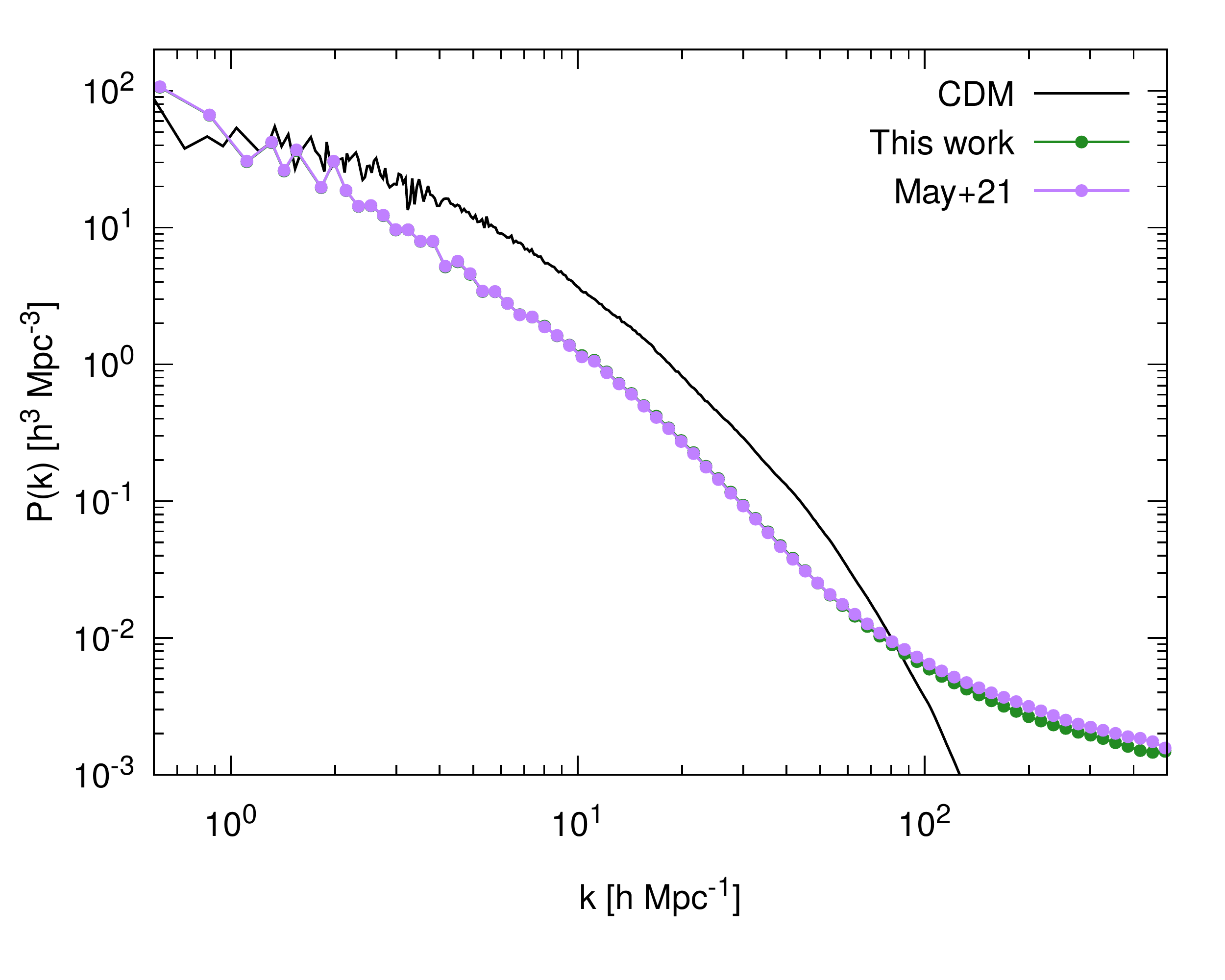}
    \caption{Comparison of the power spectrum at $z = 0$ between the code used in this work and that of \citet{Simon2021} for a cosmological test simulation.}
    \label{fig:figPk}
\end{figure}

We ran a cosmological fuzzy dark matter simulation separately with identical initial conditions generated by \textcode{MUSIC} with box size $L = \SI{10}{\Mpc\per\hHubble}$, particle mass $mc^2 = \SI{2.5e-24}{\eV}$, and number of grid cells $N^3 = 1024^3$.
The cosmological simulations are evolved until $z = 0$, and the density fluctuations are measured as the power spectrum shown in \cref{fig:figPk}.

\bsp	
\label{lastpage}
\end{document}